\documentclass[apj]{emulateapj}
\usepackage{apjfonts}
\usepackage{amsmath}
\usepackage{bmpsize}
\usepackage[colorinlistoftodos]{todonotes}
\usepackage{graphicx} 
\usepackage{url}
\shorttitle{MUSE-ings on AM1354-250}
\shortauthors{Conn et al.}
\begin{document}

\title{MUSE-ings on AM1354-250: Collisions, Shocks and Rings}

\author{Blair C. Conn\altaffilmark{1}, L.M.R. Fogarty\altaffilmark{2}, Rory Smith\altaffilmark{3,4,5} and Graeme N. Candlish\altaffilmark{5}}
\altaffiltext{1}{Gemini Observatory, Recinto AURA, Colina El Pino, Casilla 603, La Serena, Chile; bconn@gemini.edu}
\altaffiltext{2}{Sydney Institute for Astronomy, School of Physics, The University of Sydney, A28, Sydney, 2006, Australia}
\altaffiltext{3}{Yonsei University, Graduate School of Earth System Sciences-Astronomy-Atmospheric Sciences, Yonsei-ro 50, Seoul 120-749, Republic of Korea}
\altaffiltext{4}{CEA-Saclay, DSM, DAPNIA, Service d'€™Astrophysique, F-91191 Gif-sur-Yvette, France}
\altaffiltext{5}{Departamento de Astronom\'{i}a, Universidad de Concepci\'on, Casilla 160-C, Concepci\'on, Chile}

\begin{abstract}
We present MUSE observations of AM1354-250, confirming its status as a collisional ring galaxy which has recently undergone an interaction, creating its distinctive shape. We analyse the stellar and gaseous emission throughout the galaxy finding direct evidence that the gaseous ring is expanding with a velocity of $\sim$70\,km.s$^{-1}$ and that star formation is occurring primarily in HII regions associated with the ring. This star formation activity is likely triggered by the interaction. We find evidence for several excitation mechanisms in the gas, including emission consistent with shocked gas in the expanding ring and a region of LINER-like emission in the central core of the galaxy. Evidence of kinematic disturbance in both the stars and gas, possibly also triggered by the interaction, can be seen in all of the velocity maps. The ring galaxy retains weak spiral structure, strongly suggesting the progenitor galaxy was a massive spiral prior to the collision with its companion an estimated $140 \pm 12$~Myr ago.

\end{abstract}


\section{Introduction}

Collisional ring galaxies are a rare class of interacting galaxies characterised by a ring shape with a central or offset nucleus, or sometimes no identifiable nuclear region. They are thought to be the result of an almost head on collision between two objects, the target galaxy (which forms the spectacular ring) and an interloping collider, usually identifiable as a nearby galaxy.

The model for collisional ring galaxies as nearly head-on collisions was put forward by \citet{LyndsToomre1976} and \citet{TheysSpiegel1977}. \citet{LyndsToomre1976} show two simple models of head-on collisions, one with an impact parameter of zero (i.e. a centred collision) and the other slightly offset. They find that an outward propagating density wave is created in the disk of the target galaxy leading to the distinctive ring shape. Similarly \citet{TheysSpiegel1977} examine numerical models of collisions and also investigate the ultimate fate of ring galaxies, finding them to be unstable on timescales of $10^8$\,yrs. This supports the observation that these objects are relatively rare. 

The collision formation scenario can successfully reproduce the observed geometry of individual ring galaxies. \citet{MihosHernquist1994} reproduce the geometry of the Cartwheel using a head-on collision with an impact parameter of zero. \citet{LyndsToomre1976} find that to reproduce the geometry of II Hz 4, a system in which the collision creates not one but two rings, a modestly off-centre collision fares better. In fact a range of angles of incidence and impact parameters are investigated in \citet{HuangStewart1988}, where the authors find that impact angles up to 30$^{\circ}$ can produce rings with a central nucleus, though larger angles of impact coupled with a non-zero impact parameter can dislodge the nucleus leading to an empty ring. Although a tight impact parameter produces the best rings, values up to half of the disk radius still produce arc-like structures. 

Apart from their distinct morphology, ring galaxies also share some other observational properties. Studying a sample of 11 ring galaxies \citet{AppletonMarston1997} found they displayed blue optical and IR colors in the ring. They also found, for the four biggest rings in their sample, a sharp radial color gradient. The rings show red interior colors with bluer colors in the ring itself. For two of these galaxies the color was shown to become redder again beyond the ring radius. These results are consistent with the scenario of an outward propagating ring sweeping up gaseous material and triggering star formation. In the large rings with an outer red region the ring may not have propagated to the edge of the original disk yet.

Ring galaxies also show enhanced star formation rate (SFR) when compared to `normal' disk galaxies. \citet{AppletonStruckMarcell1987} found enhanced far-infrared luminosity and infrared to B-band luminosity ratio, which they attributed to an enhancement in star formation. Using the same sample as \citet{AppletonMarston1997}, \citet{MarstonAppleton1995} found that ring galaxies display H$\alpha$ luminosities comparable to those found in starburst galaxies.

Kinematic studies can also shed light on the formation of collisional ring galaxies. Early examples of this work include \citet{TheysSpiegel1976} who found signatures of an expanding ring in VII Zw 466 and \citet{Fosbury1977} who examined the kinematics of the Cartwheel galaxy. These kinematic properties can be very useful inputs to simulations, helping to understand the dynamics of the collisions. Detailed spectroscopic data from Auriga's wheel \citep{Conn2011} allowed \citet{Smith2012} to reproduce the system very effectively, constraining the timescale and mass ratio for the collision and even predicting the coalescence of the system into one galaxy on timescale of $\sim400$ Myr from its current configuration.

This is clearly an area where integral field spectroscopy (IFS) can contribute significantly, constraining the dynamics of the collisional ring not just along a few carefully chosen axes but in a spatially resolved fashion. \citet{Fogarty2011} demonstrated this with an IFS study of Arp 147. The advent of the Multi Unit Spectroscopic Explorer (MUSE) on the ESO/VLT\footnote{European Southern Observatory/Very Large Telescope} is an excellent opportunity to study the dynamics and other physical properties of these systems in unprecedented detail.

\section{AM1354-250: A Collisional (Double) Ring Galaxy}
\begin{figure*}
	\centering
	\includegraphics[width=\textwidth]{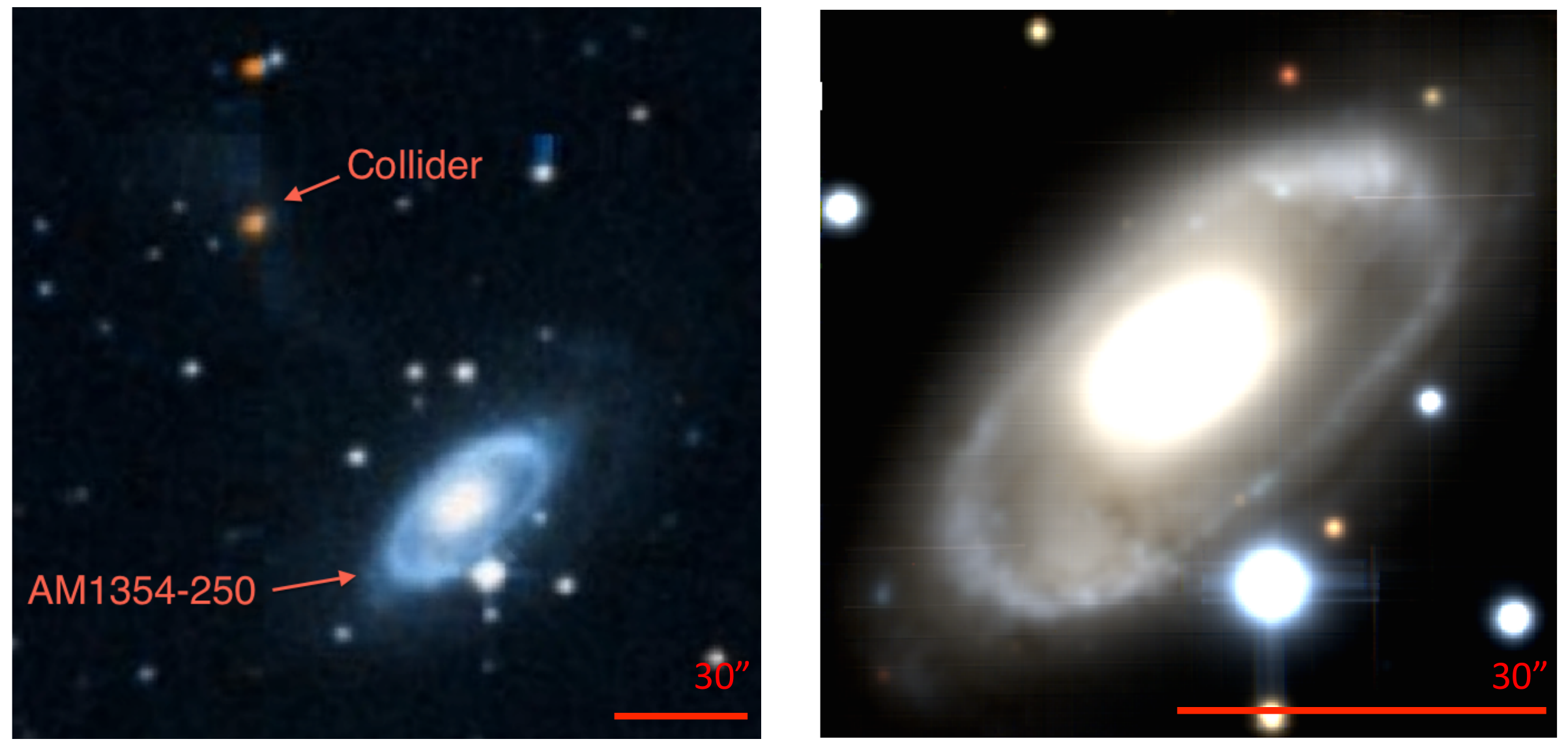}
	\caption{\label{fig:dss_muse} Left panel: HiPS color Digitized Sky Survey image\footnote{The red component has been built from POSS-II F, AAO-SES,SR and SERC-ER plates. The blue component has been build from POSS-II J and SERC-J,EJ. The green component is based on the mean of other components. http://archive.stsci.edu/dss/acknowledging.html}, showing AM1354-250 and its colliding galaxy. The ring of triggered star formation is clearly visible along with the diffuse outer disk and wispy tidal debris leading to the red core of the colliding galaxy. North is up and East is to the left. Right panel: False color RGB image generated from the reduced MUSE data cube by applying masks equivalent to SDSS\footnote{Sloan Digital Sky Survey} $g$, $r$ and $i$ bands (R$=i$, G$=r$, B$=g$). The color image has then been assembled using DS9\footnote{http://ds9.si.edu/site/Home.html}. The 0.2 by 0.2 arcsecond spaxels provide the highest resolution image of this galaxy to date. Easily visible are the dust lanes and remnant spiral structure of the galaxy. The disk appears to be warped. North is up and East is to the left.}
\end{figure*}

AM1354-250 is an archetypal collisional ring galaxy from the catalogue of \citet{Madore2009}, based on the earlier \citet{ArpMadore1987} catalogue of peculiar galaxies. It has a strong visible star forming ring and the collider galaxy (2MASX J13571798-2513291) can be found close-by with stellar tidal debris linking the two. As discussed by \citet{Madore2009}, it is also a potential member of a rare subclass of collisional ring galaxies, those with double rings. The putative second ring is thought to be the diffuse stellar material that can be seen in the left panel of Figure~\ref{fig:dss_muse} at much greater radius than the dominant ring. However, it is unclear whether this material can truly be considered another ring or is simply the low surface brightness extension of the disk. 

The collisional ring galaxy has a redshift of 0.02068 \citep{2008AJ....135.2424O} and is located in the constellation of Hydra (13h57m13.9s -25d14m45s J2000.0). Its R-band luminosity reaches 12.06$\pm 0.09$ magnitudes with a B-band of 13.70$\pm 0.09$ magnitudes. It has significant flux in the infrared with J-, H- and K-bands between 10.265 and 9.285 magnitudes. The left-hand panel of Figure~\ref{fig:dss_muse} shows a near perfect ring of star formation with a radius of approximately 26 arcseconds ($\sim 10$ kpc). A disturbed nearby companion is the most likely collider and has a redshift of 0.02043. It can be seen near the top-left corner of the left-hand panel of Figure~\ref{fig:dss_muse} and appears to be associated with debris from its interaction with AM1354-250. Table~\ref{datatable} presents the basic information for AM1354-250 and its companion. 

Our knowledge of this galaxy system is limited to various instances in which it appears in other major surveys (e.g. 6dF from \citet{2004MNRAS.355..747J} and 2MASS with \citet{2006AJ....131.1163S}) and a dedicated study of this object by \citet{Wallin1995}. In this latter study the author analyses UBVRI photometry to show that AM1354-250 has a weak burst of star formation caused by a collision with its companion. The author also uses simple stellar population fitting to estimate that the collision occurred less than 100 Myr ago. Through numerical modelling of the ring \citet{Wallin1995} determines that the star-forming ring is expanding with a velocity of $\sim$95 km.s$^{-1}$. The author also demonstrates that the expanding ring is almost exclusively associated with a young stellar population and is largely absent in old underlying stellar population of the galaxy. There is no discussion on the presence of a double ring in \citet{Wallin1995} as their analysis does not extend to large enough galactic radii to sample the outer disk. Their dataset also does not seem deep enough to detect the faint stellar material discussed in \citet{Madore2009}.

In \citet{Hibbard2001}, the HI gas distribution of AM1354-250 can be seen in their Figure~95\footnote{\url{https://www.nrao.edu/astrores/HIrogues/webGallery/RoguesGallery12.html}}. The HI is clearly heavily disrupted, forming two lobes of extended emission along the major axis of the galaxy perpendicular to the line joining the galaxy with the collider. These lobes are visible out to distances three times that of the visible stellar ring ($\sim$75 arcseconds radially) and the centre of the galaxy is apparently devoid of HI. There is also no detection of any HI belonging to the collider galaxy or connecting the two galaxies. 

AM1324-250 is an ideal candidate for IFS observations with MUSE. It is a bright, nearby system with a ring diameter of 53 arcseconds \citep{Madore2009}, ideally matched to the MUSE field of view. The outer second ring is mostly beyond the MUSE field of view but the MUSE observations can shed light on the formation of this ring galaxy and provide constraints and insight into the formation of ring galaxies generally. 
 
\begin{table}
\centering
\begin{tabular}{|l|c|c|}
\hline
& AM1354-250 & 2MASX \\
& & J13571798-2513291 \\\hline
RA (J2000) & 13:57:13.8 & 13:57:18.0  \\ 
DEC (J2000) & -25:14:44.9 & -25:13:29.3\\
Redshift\footnote{NASA Extragalactic Database} & 0.020681$\pm$0.000060& 0.020434$\pm$0.000103\\
Image scale $(kpc/")$\footnote{This work\label{MUSE}}& 0.437 & \nodata\\
Radius (") & 26.5\footnote{\citet{Madore2009}\label{Madore}} & 7.5$^{\ref{MUSE}}$\\
Radius (kpc) & 11.6$^{\ref{Madore}}$ & 3.2$^{\ref{MUSE}}$\\
B Mag & 13.68$\pm$0.21 & \nodata\\
R Mag & 12.06$\pm$0.09 & \nodata\\
Ks Mag & 9.285$\pm$0.025 & 12.727$\pm$0.139 \\
E(B-V)\footnote{\citet{SF2011}}$^{,}$\footnote{http://irsa.ipac.caltech.edu/applications/DUST/} & 0.0724 & 0.0716\\
A$_Ks$ (mag) & 0.026 & 0.026 \\
Distance\footnote{Luminosity distance, calculated using WMAP9 cosmological parameters, \citet{Hinshaw2013} with H$_0=69.32$, $\Omega_{\Lambda}=0.714$ and $\Omega_{M}=0.286$.} (Mpc) & 90.882 & 89.780\\
Stellar Mass\footnote{Only for the bulge of AM1354-250} & $\sim$1.5$\times$10$^{11}$M$_{\odot}$& $\sim$1.3$\times$$10^{10}$M$_{\odot}$\\\hline
 
\end{tabular}
\caption{\label{datatable} AM1354-250 and 2MASX J13571798-2513291: Information compiled from the literature and this study.}
\end{table}

\section{Data Reduction and Analysis}
\label{sec:Data}
MUSE is a second generation instrument located at the Nasmyth focus of UT4 of ESO VLT at Cerro Paranal. It is an adaptive optics (AO) assisted wide field integral field spectrograph. It has a field of view (fov) of 1 square arcminute, with the fov consisting of 24 identical Integral Field Units (IFUs) that provide contiguous spatial coverage. Each IFU provides spatial sampling of 0.2$\times$0.2 arcseconds. The resultant spectra cover a wide wavelength range from 0.465-0.93$\mu$m with a resolution of 2000 (at 0.465$\mu$m) to 4000 (at 0.93$\mu$m). This provides for an instrumental velocity resolution of $\sim$150 km.s$^{-1}$ at the blue end and $\sim$75 km.s$^{-1}$ at the red end.

The observations presented here were taken as part of the ESO/MUSE Science Verification (SV) observations under the ESO Programme ID 60.A-9310(A). The programme was completed on June 29, 2014 and consists of 3 science observations of 1200 seconds dithered on target with 3 interleaved sky frames of 120s each. The instrument used the Wide Field Mode with no AO and had the `Blue' filter setting which provides a wavelength coverage of $0.48-0.93\mu$m. We use the calibration frames (biases, lamp flats and darks) taken throughout the SV time, with the arc and lamp flat frames observed as attached calibrations taken with the science data. We also use standard star, twilight and astrometry field frames observed on June 29 2014.

\begin{figure*}
\centering
\includegraphics[width=1.0\textwidth]{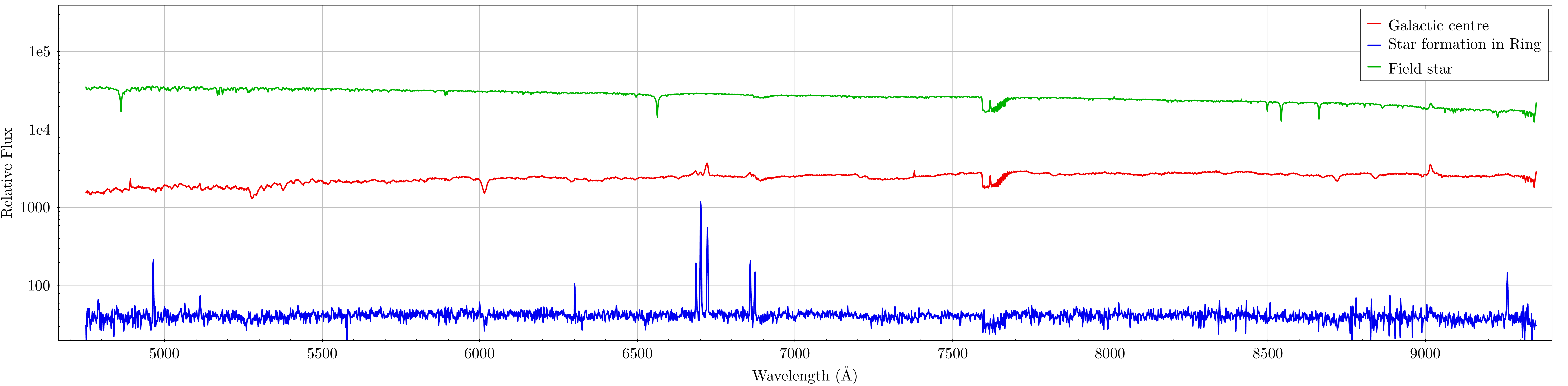}
\caption{\label{fig:spectra}Examples of spectra obtained from our dataset. The top (green) spectrum is from the brightest star in the field, the middle (red) spectrum is from the core of AM1354-250 and the lower (blue) spectrum is from a large star-forming region in the ring. In the middle and lower spectra, the presence of emission lines are clearly visible along with the underlying stellar component.}
\end{figure*}

The data are reduced using the MUSE pipeline Version 1.0 through the {\tt esorex} interface. We follow the procedure outlined in the MUSE data reduction cookbook \footnote{ftp://ftp.eso.org/pub/dfs/pipelines/muse/muse-pipeline-cookbook-1.0.pdf}. We first create a master bias and master dark frame, combining 22 and 5 individual observations respectively. Next a master flat field is created using lamp flat data observed over several days. The master flat is used to trace the edges of the individual slits in each MUSE IFU, generating trace tables for each unit. A wavelength solution is created using a set of arc frames taken in attached calibrations with our data (9 frames in total, 3 each using the Ne, Xe and HgCd arc lamps). Next, an accurate line spread function (LSF) is generated using a much larger set of arc frames taken throughout the course of the SV time. A large number of files is needed to accurately measure the faint wings of the emission lines to characterise the LSF. We used 14 exposures for each lamp in this step. Lastly, the twilight sky observations are processed to assess the illumination correction for each IFU.

The calibration frames are then used, along with the static calibrations provided by the MUSE instrument team, to remove the instrument signature from each of the target, sky, standard star and astrometry field observations. This step creates the necessary pixel tables to create data cubes for each frame. The sky and standard star pixel tables are used to generate the necessary files to sky subtract and flux calibrate the data respectively. In the penultimate processing step, an astrometric solution is generated to ensure that the sky coordinate system in the MUSE data cubes is correct. Finally three fully-calibrated data cubes are created from the pixel tables for each of the three 1200s target exposures. Images generated from these cubes are used to manually measure offsets between the frames, which are finally combined into one master data cube. In this latter step, three images are also formed from the resulting cube, using the SDSS $g$, $r$ and $i$ filter curves \citep{2010AJ....139.1628D}. An RGB composite using these images is shown in the right-hand panel of Figure \ref{fig:dss_muse}. The detail in this spectacular image shows the power of MUSE, especially when one remembers that each spaxel in the image also contains spectral information.	An example of the spectra obtained in this dataset, produced using {\sc topcat}\footnote{\url{http://www.starlink.ac.uk/topcat/}} \citep{Taylor2005}, can be seen in Figure~\ref{fig:spectra}. The galaxy data has many features in both absorption and emission, we report on the properties of several of the most dominant features.

\subsection{Isophotal Analysis and Mass Estimate}
\begin{figure}
\centering
\includegraphics[width=0.49\textwidth]{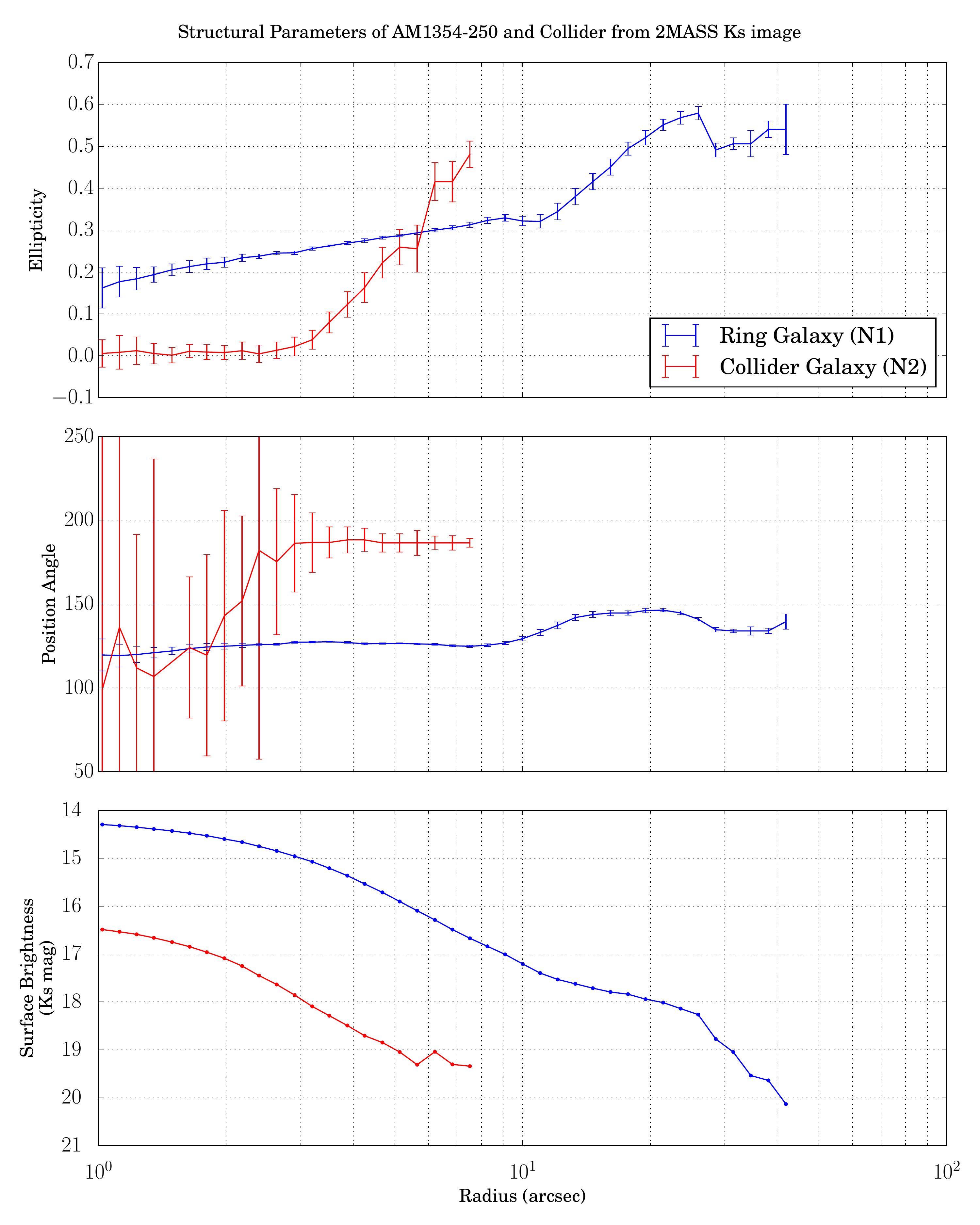}
\caption{\label{fig:struct_params}Isophotal analysis results of the 2MASS Ks-band image (1 arcsecond per pixel resolution) from {\sc ellipse} IRAF routine. All parameters are plotted against radius in arcseconds. The top panel shows how the ellipticity varies with radius for the nuclei of both galaxies. The second panel shows the position angle of the isophotal ellipses (for ellipses with errors less than 20 degrees) and the third panel shows the surface brightness in magnitudes per square arcsecond. The ring galaxy isophotes (in blue) extend approximately 40 arcseconds from the central isophote, while the collider galaxy isophotes (in red) stop around 10 arcseconds from the central pixel. The error bars are obtained from the output of the IRAF routine.}
\end{figure}

We use the Ks-band image from 2MASS \citep{2006AJ....131.1163S} as the basis for an isophotal analysis of both the ring galaxy and the collider galaxy. The image has been sky subtracted and the point sources removed. While some residuals from the foreground stars are still visible these do not overlap with our isophotes and no masking has been employed prior to analysis. We use the IRAF package {\sc ellipse} to derive radial profiles of surface brightness, ellipticity and position angle (PA) of the fitted ellipses for the ring galaxy and the collider. The results are shown in Figure~\ref{fig:struct_params}, with the ring shown in blue and the collider in red. The {\sc ellipse} routine uses the brightest pixel within the nucleus as the centre and is forced to avoid isophotes within the central 1 arcsecond.

The bulge of AM1354-250 shows an ellipticity slowly increasing with galactocentric radius. A rapid increase in ellipticity occurs at the radius of the unobscured disk ($\sim10$ arcseconds) and the ellipticity continues to increase until the last few isophotes. A similar radial variation in PA of is also seen, with the inner bulge having a PA around 125$^{\circ}$, compared to the inner disk (dominated by the spiral arms) at a PA of 145$^{\circ}$, and the ring with a PA of $\sim$135$^{\circ}$. The suspected outer ring is not visible in the 2MASS image and so we have no measurements of its properties from this analysis. 

By comparison, the collider galaxy is much smaller and rounder with the fitted isophotes reaching the sky background within 10 arcseconds of the core. The very low ellipticity of the galaxy results in a large error in the PA and so reliable PA estimates are only found beyond a radius of 3 arcseconds. The PA in this inner region is $\sim$186$^{\circ}$ and corresponds approximately to the position of the tidal debris that can be clearly seen in the left-hand panel of Figure~\ref{fig:dss_muse}.

To estimate the mass of these galaxies we use the Ks-band Luminosity to Mass scaling relation found in \citet{2003ApJS..149..289B}. For the ring galaxy we limit the conversion to only include the central bulge region while for the collider we take the flux enclosed in the final isophote. In both cases, this is within 10 arcseconds of the central pixel. The scaling relation has the form,

\begin{equation}\label{eqn:mass}
   \left(\frac{M}{L}\right)_{\odot} \sim 0.95\\
\end{equation}

From this, we estimate the bulge in the ring galaxy to have a minimum mass of $\sim$1.5$\times$10$^{11}$M$_{\odot}$ and the collider galaxy to be $\sim$1.3$\times$$10^{10}$M$_{\odot}$, this relation has a 30\% error. 

\begin{figure}[b]
\centering
\includegraphics[width=0.49\textwidth]{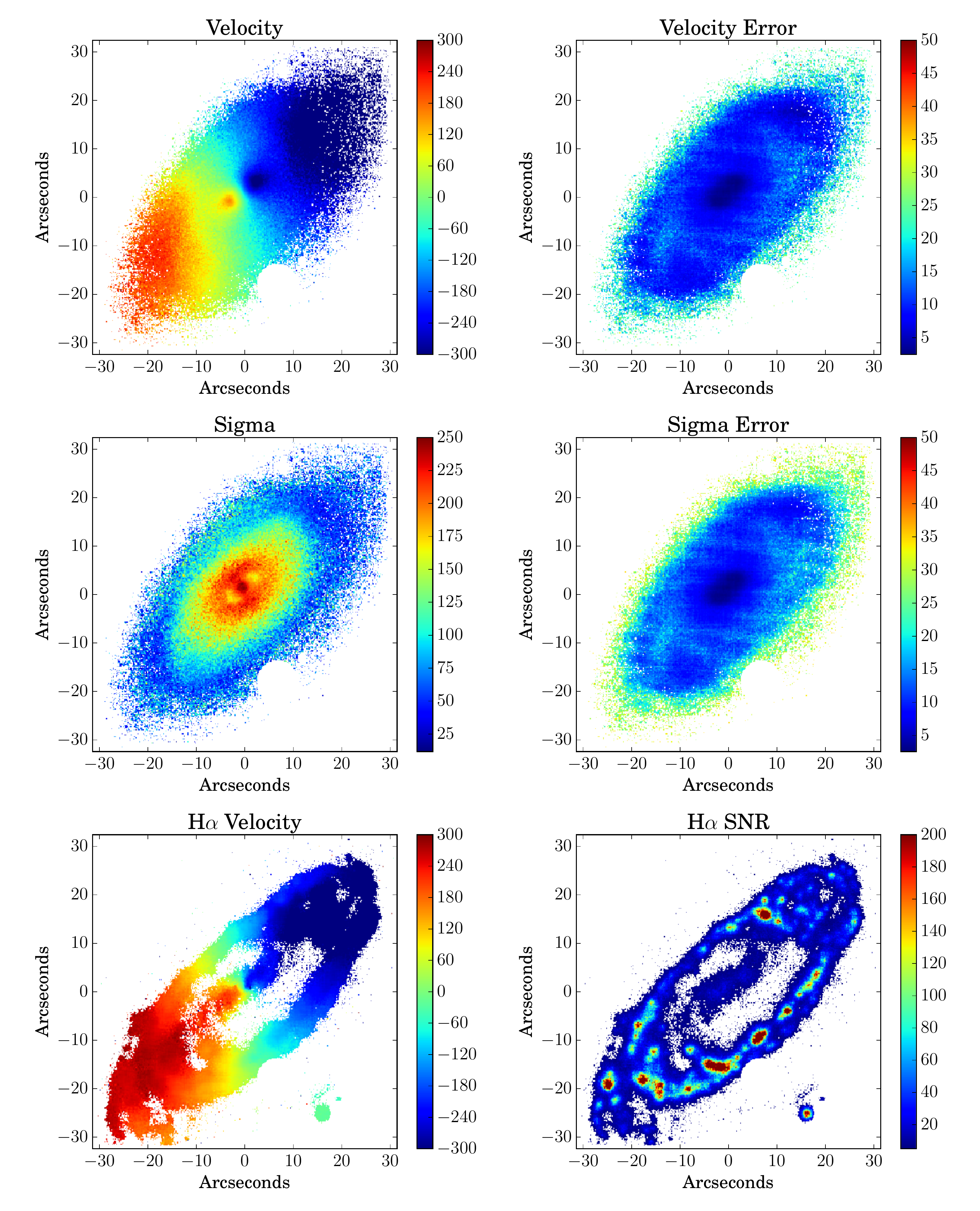}
\caption{\label{fig:stellar_kin}Properties extracted from pPXF fits to the continuum. The Left column shows the velocity map and the velocity dispersion for the stellar continuum fits with the H$\alpha$ velocity map at the bottom for comparison. The Right column shows the respective errors for the continuum fit and the Signal to Noise ratio for the H$\alpha$ line at the bottom. The minimum S/N ratio is 5 and the velocities are in km.s$^{-1}$. }
\end{figure}

\subsection{Analysis of MUSE Data}
The MUSE data cube is analysed in two steps. First, the stellar continuum is fit using the penalised pixel fitting (pPXF) routine of \citet{Cappellari04}. A central spectrum is extracted from the bulge region of the data cube and fit using the 985 MILES stellar templates \citep{SanchezBlazquez06} and an additive 5th order polynomial. The weighted and combined best fit template is then used to fit all individual spaxels in the data cube, again in combination with an additive 5th order polynomial. The algorithm attempts to fit all spaxels in the data cube, returning null values only where the fit fails. For each spaxel in the MUSE data cube this produces a best fit template spectrum and a line of sight velocity distribution (LOSVD), parametrised as a Gaussian. The resulting velocity and velocity dispersion maps for AM1324-250 are presented in Figure~\ref{fig:stellar_kin}.

\begin{figure*}
	\centering
	\includegraphics[width=\textwidth]{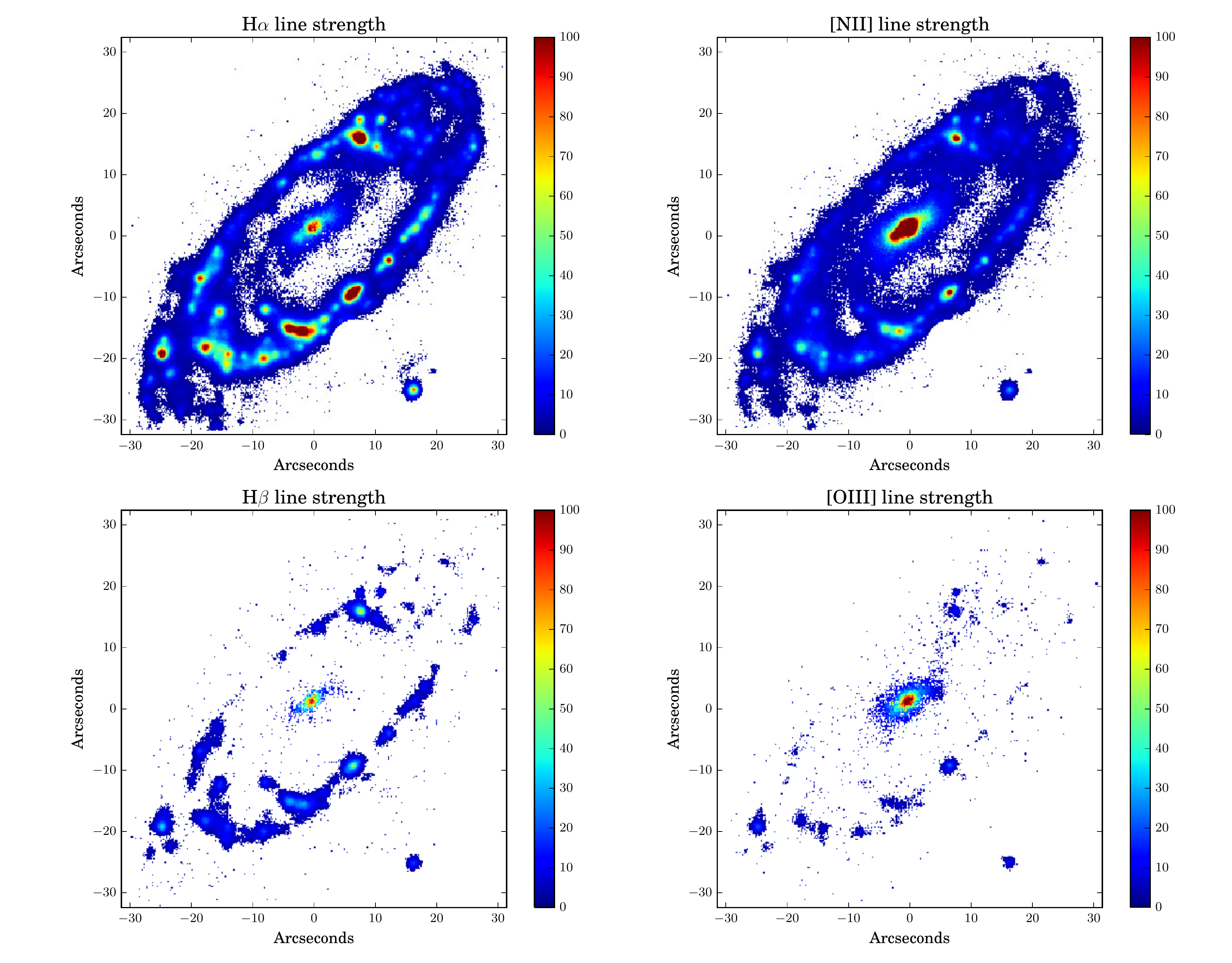}
	\caption{\label{fig:gas_linemaps}
Line maps extracted from Gaussian profile fits to the H$\beta$, [OIII]$\lambda$5007, H$\alpha$ and [NII]$\lambda$6583 emission lines. The line maps do not have absolute flux calibration and so the relative flux levels have been scaled linearly from 0 to 100. To improve the contrast we have let some values saturate allowing the fainter structures to become visible. The galaxy has clear remnants of a two arm spiral structure along with the dominant star forming ring. The disk is extends beyond the ring and is possibly warped. Interestingly, there is a very bright emission line source in the lower right of the image, quite far from the disk. On the scale of this figure it is approximately 2 kpc across.}
\end{figure*}

\begin{figure*}
	\centering
	\includegraphics[width=1.0\textwidth]{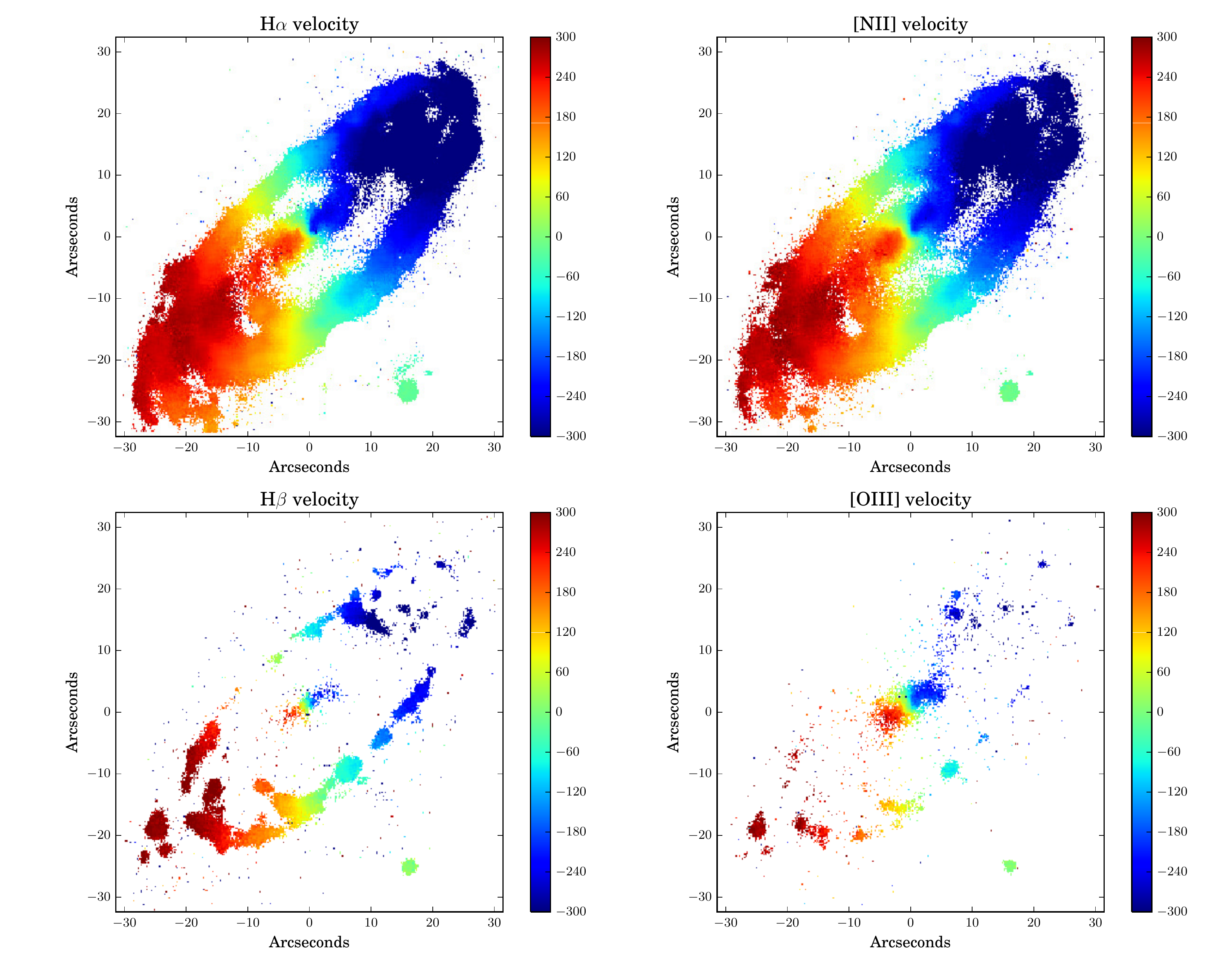}
	\caption{\label{fig:gas_emission}
Velocity maps extracted from Gaussian profile fits to the H$\beta$, [OIII]$\lambda$5007, H$\alpha$ and [NII]$\lambda$6583 emission lines. The line maps show the clear signature of rotation in all lines including the inner component near the core which is decoupled from the disc. The star forming region beyond the disk has velocities roughly consistent with the disk.}
\end{figure*}

Second, the best fit to the continuum in each individual spaxel is subtracted from that spaxel to produce an emission line data cube. Once the emission line cube is created, four strong emission lines (H$\beta$, [OIII]$\lambda$5007, H$\alpha$ and [NII]$\lambda$6583) are fit with single Gaussian profiles. The S/N of each line is assessed individually comparing the peak emission to noise in neighbouring continuum-subtracted bands and only spaxels with S/N$\ge 5$ are fit. H$\alpha$ is the strongest emission line in our dataset and to illustrate the typical S/N distribution in the emission line maps, the H$\alpha$ velocity and S/N maps are also shown in the bottom two panels of Figure~\ref{fig:stellar_kin}. For ease of comparison, all the emission line flux maps are presented in Figure~\ref{fig:gas_linemaps} with the corresponding velocity maps shown in Figure~\ref{fig:gas_emission}.

\section{Results}

\subsection{Stellar Kinematics and Gas Properties of AM1354-250}
\label{sec:kinematics}
The stellar kinematics for AM1354-250 are shown in Figure \ref{fig:stellar_kin}, with the velocity in the top panel and the velocity dispersion in the bottom. From both maps it is clear that there are two separate rotation-supported kinematic components. The inner disk, which peaks at a velocity of $\sim225$\,km.s$^{-1}$ at a radius of $\sim3.5$ arcsecs, is superimposed on the outer disk and ring structure. This inner stellar disk is kinematically distinct but aligned with the main disk. It can be seen in the velocity dispersion map as two depressions either side of the central peak. 

The core of the galaxy has some of the strongest emission present in all emission lines (Figure~\ref{fig:gas_linemaps}) and internal to the star forming ring there are regions with little gas (perhaps corresponding to the inter-arm regions in the original spiral galaxy, before the collision occurred). At the location of the ring there is a clear abundance of ionised gas as illustrated by the two strongest lines, H$\alpha$ and [NII]. Beyond the ring, the gas disk extends to large galactic radii, primarily along the major axis as seen from our line of sight. There is also a large star forming region present in the lower right corner of the maps that may be related to the faint stellar disk seen in the left panel of Figure~\ref{fig:dss_muse}. The gas kinematic maps (Figure~\ref{fig:gas_emission}) do not show a central disk in the same manner as the stellar maps, but rather indicate a more continuously rising rotation curve.

We use {\sc Kinemetry} \citep{Krajnovic2006} to further analyse the stellar and H$\alpha$ velocity maps. {\sc Kinemetry} decomposes a given map into a series of concentric ellipses and fits a harmonic modal expansion to each ellipse. For the velocity maps, the assumption is that the profile along any given ellipse follows a cosine law, encoded in the first term of the expansion (with coefficient $k_{1}$). Therefore, any deviation from rotation will result in fitting non-zero values to the higher order terms in the expansion (with coefficients $k_{3}$ and $k_{5}$). The ratio of $k_{5}$/$k_{1}$ therefore gives an indication of whether a galaxy contains significant deviation from simple rotation. \citet{Krajnovic2011} use a cut-off of $k_{5}/k_{1} < 0.04$ to indicate that a galaxy is well-described as a regularly rotating. Extending this finding from early-type to late-type galaxies, \citet{2012A&A...542A..54B} confirm that this relation works well for undisturbed spirals. 

\begin{figure}
\centering
\includegraphics[width=0.45\textwidth]{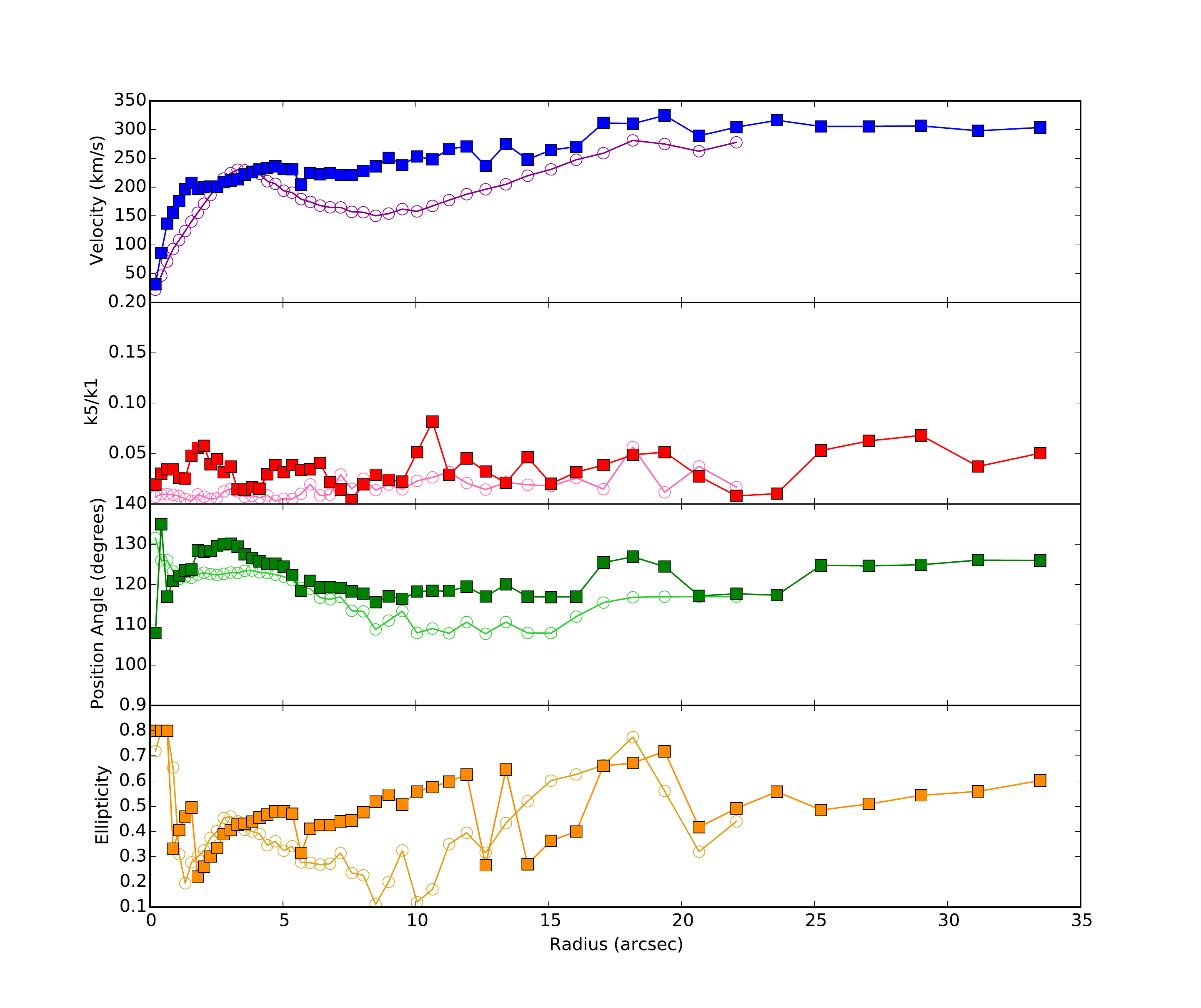}
\caption{\label{fig:kinemetry}Results from {\sc kinemetry} for the stellar and gas velocity fields of AM1354-250. The open round points indicate values for the stellar velocity map and filled square points for the gas.}
\end{figure}

The kinemetric output for both the stellar and H$\alpha$ velocity maps of AM1354-250 are shown in Figure \ref{fig:kinemetry}. The first panel shows the rotation curves derived from both the stellar and H$\alpha$ velocity maps. The central, kinematically distinct, stellar disk is clearly seen. The H$\alpha$ velocity profile rises steadily and shows no kinematically distinct component in the centre.

The second panel shows $k_{5}/k_{1}$. The stellar component displays regular rotation across the entire disc. The H$\alpha$ profile displays several regions where $k_{5}/k_{1} > 0.04$ suggesting non-regular rotation. The first is in the very centre of the galaxy ($\sim$2 arcseconds). The second region is around 10 arcseconds, and this could be related to kinematic disturbance due to the spiral arms. The third region is a broad excess peaking around $\sim$28 arcseconds. The star-forming ring looks warped at this radius, so this could cause the excess power in $k_{5}$. In general however, the galaxy is very well behaved and does not resemble the kinds of departures from regular rotation that \citet{2012A&A...542A..54B} describe as a disturbed system. 

The third and fourth panels of Figure~\ref{fig:kinemetry} show the measured position angle and ellipticity from {\sc Kinemetry}. The values are consistent with those presented in Figure~\ref{fig:struct_params} from the 2MASS Ks-Band image. 

\subsection{Ionisation Mechanisms and Star Formation}
\begin{figure}[b]
\centering
\includegraphics[width=0.45\textwidth]{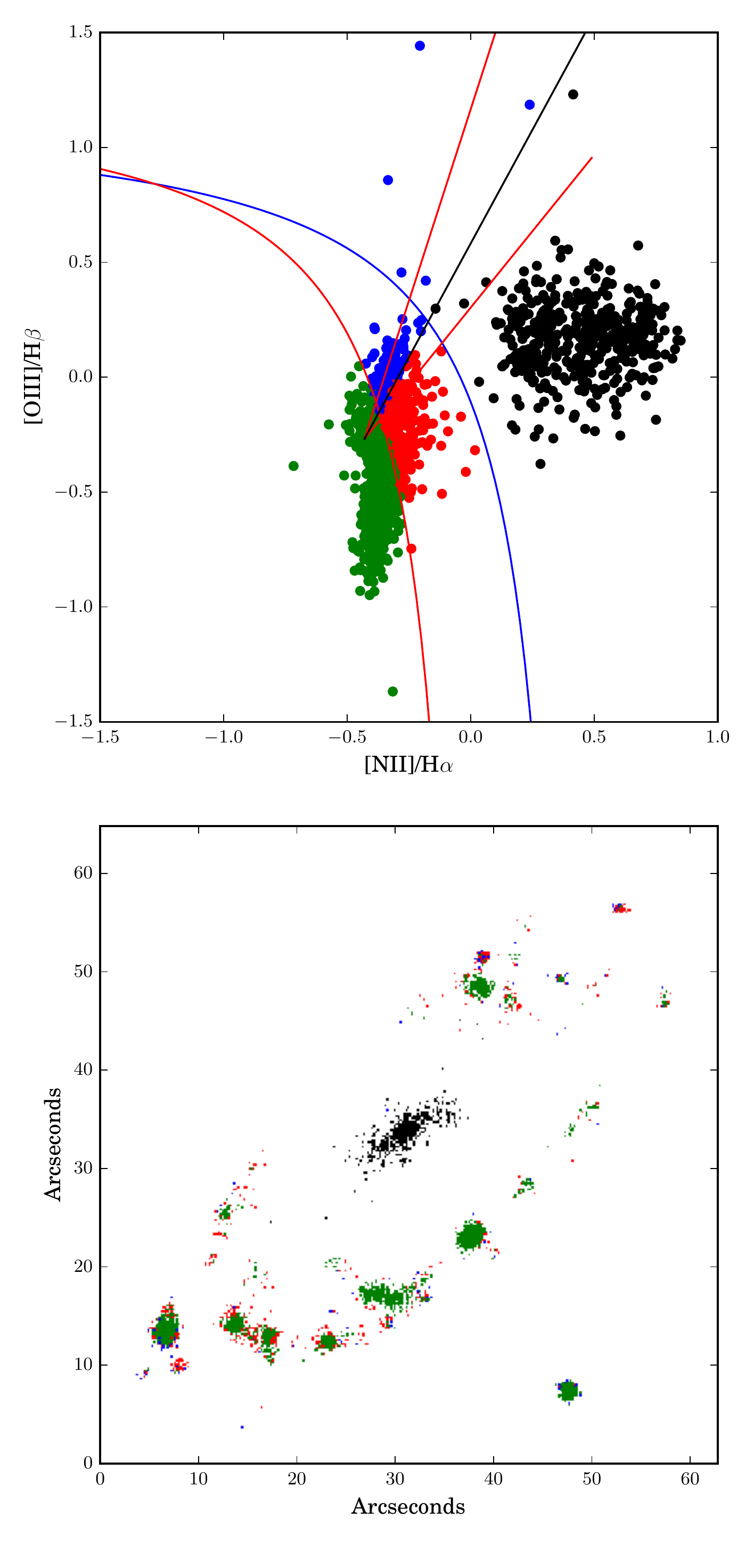}
\caption{\label{fig:bpt}Top panel: The IDD for the [OIII]/H$\beta$ and [NII]/H$\alpha$ line ratios. Each point corresponds to a single MUSE spaxel and the points are color-coded according to the dominant ionisation mechanism giving rise to the spectrum at that point. Bottom panel: The spatial distribution of points in the top panel are shown, with the same color-coding: Black points - LINER-like emission; Green points -  Star formation; Blue and Red points - shocked gas emission.}
\end{figure}

The gas in AM1354-250 shows different ionisation properties in different parts of the galaxy. To probe the ionised gas we examine the four strongest emission lines in the MUSE spectra. We calculate the ratio of [OIII]/H$\beta$ and [NII]/H$\alpha$ in each spectrum where all four lines have S/N$ \ge 5.0$. This excludes many spectra where the [OIII] and/or H$\beta$ emission lines are weak. The top panel of Figure \ref{fig:bpt} shows the dominant ionisation mechanisms in a BPT diagram \citep{1981PASP...93....5B}. The limiting starburst line \citep{Kewley2001} is shown in blue with the empirical star formation line in red \citep{Kauffmann2003}. The black line gives the division between AGN (Active Galactic Nuclei) and LINER-like (Low-Ionization Nuclear Emission-line Regions) emission, this is an empirical division from \citet{SBH2010}. The points represent single MUSE spaxels and are color-coded according to position on the BPT diagram. The bottom panel of Figure \ref{fig:bpt} shows the spatial distribution of these points. The black points, with LINER-like emission, all fall in the centre of the bulge, perhaps indicating a weak central AGN, or shocked gas from an outflow. The green points, indicative of ionisation from star formation, are clustered around the star-forming knots in the ring. Some gas in the ring also shows indications of being shocked (the blue and red points). This is also consistent with the scenario of an outward expanding ring which sweeps up and shocks gas as it triggers star formation.

\begin{figure}
\centering
\includegraphics[width=0.45\textwidth]{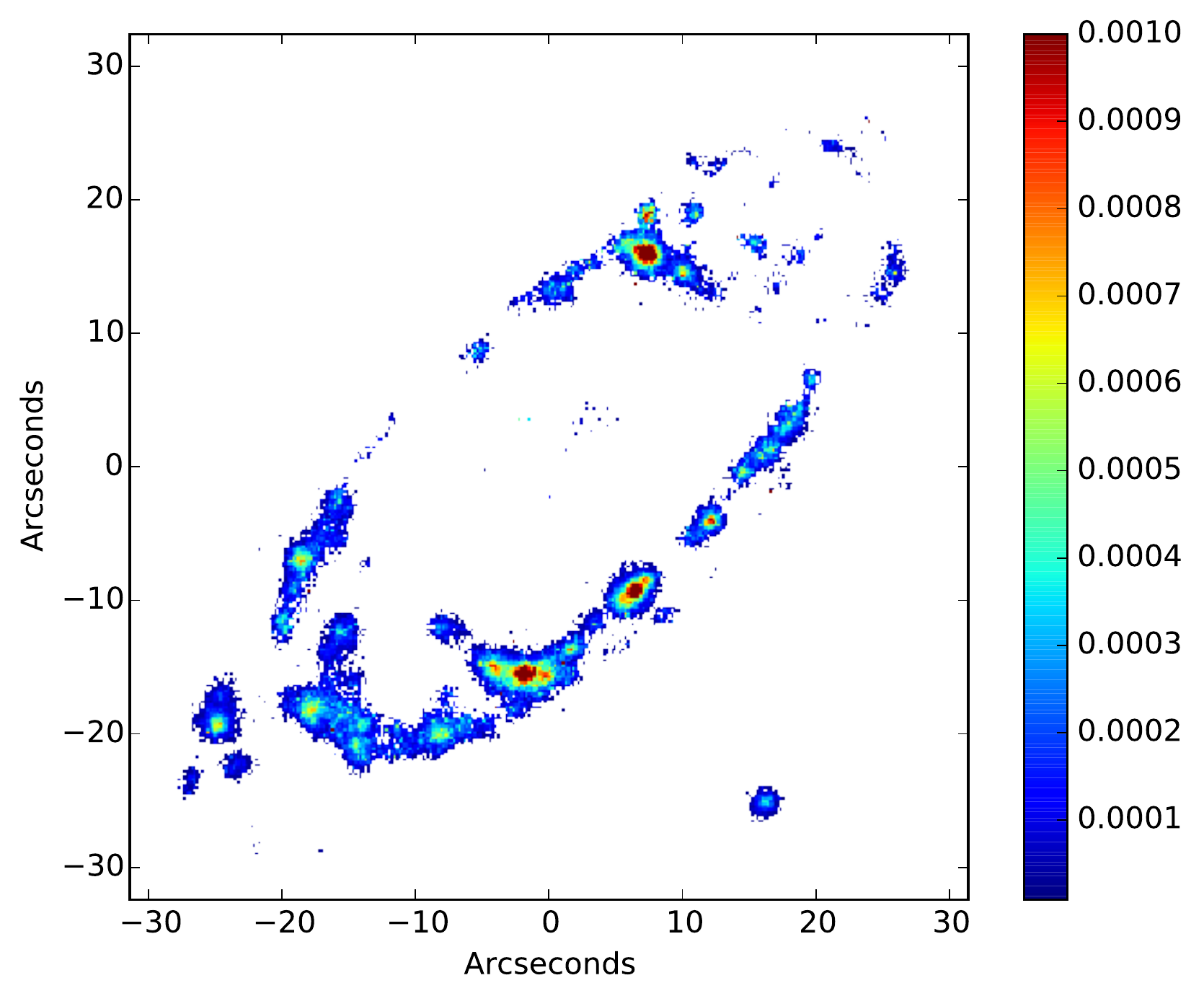}
\caption{\label{fig:sfr}A map of star formation in AM1354-250. The color bar is in units of solar masses per year.}
\end{figure}

\subsection{Collision Parameters}

We extract various parameters of the interacting galaxies by combining our MUSE data with 2MASS data. In particular, we are interested in determining the physical separation of the two galaxies and their relative velocities, the rotation and expansion velocities of the gas in the ring galaxy and to make an estimate of the mass of the system. Finally, we will attempt to determine the age of the collision itself.

To determine distances we assume a distance to the ring galaxy of $90.882$~Mpc, as given in Table \ref{datatable}, implying $0.437$~kpc/arcsec for a WMAP9 cosmology. From the false color RGB MUSE image (Figure~\ref{fig:dss_muse}) we determine the ring diameter along the semi-major axis to be $20.0 \pm 1.6$~kpc. The ring inclination is measured to be $63 \pm 3$ degrees, assuming a circular ring, accounting for uncertainty on the location of the ring edge. The line-of-sight separation between the ring galaxy and the collider is measured to be $41.3 \pm 0.7$~kpc. Accounting for inclination, this corresponds to a true separation of $46.4 \pm 2.1$~kpc. The collider is likely to be well within the virial radius of the spiral galaxy.

The relative velocities of the two galaxies can be found from the redshift difference between the collider and the ring. Using the values from Table \ref{datatable}, $\Delta$z is $2.47 \times 10^{-4}$, which gives a LOS velocity difference of $74$~km.s$^{-1}$. This implies a true velocity difference of $166 \pm 16$~km.s$^{-1}$ when accounting for inclination.

Using the MUSE velocity maps (Figure~\ref{fig:gas_emission}), we extract both the expansion velocity of the ring and the LOS rotation velocity of the disk. To measure the LOS expansion velocity of the ring we sample the point where the disk semi-minor axis meets the ring. This avoids the effect of the ring rotation contaminating the expansion velocity measurement. We find a value of $\sim$60\,km.s$^{-1}$. The LOS rotation velocity is measured where the disk semi-major axis meets the ring, which avoids the effect of the ring expansion, and is determined to be $\sim$275\,km.s$^{-1}$. Accounting for the inclination this corresponds to a true expansion velocity of $\sim$70\,km.s$^{-1}$ and a true rotation velocity of $\sim$318\,km.s$^{-1}$. This confirms the results from our kinemetric analysis in Section \ref{sec:kinematics}, which showed little deviation from a rotating disk. 

If we assume the pre-collision spiral galaxy has a circular velocity of $\sim$320\,km.s$^{-1}$, this suggests the original spiral galaxy has a total mass of approximately $10^{13}$~$M_{\odot}$. A halo of this mass is expected to have a virial radius of $\sim$400~kpc in the standard cosmological model \citep{2001MNRAS.321..155L}. As the collider and ring galaxy are separated by only $\sim$40~kpc, the collider can be considered to be still deep within the potential well of the spiral, and so probably it is being seen shortly after the collision. The fact that the true velocity difference between the collider and spiral is roughly half that of the circular velocity of the spiral also implies that the collider was probably bound to the spiral before the collision (i.e. a satellite of the spiral), unless dynamical friction has substantially slowed the collider during the collision. 

\begin{figure*}
\centering
\includegraphics[angle=0,width=0.9\textwidth]{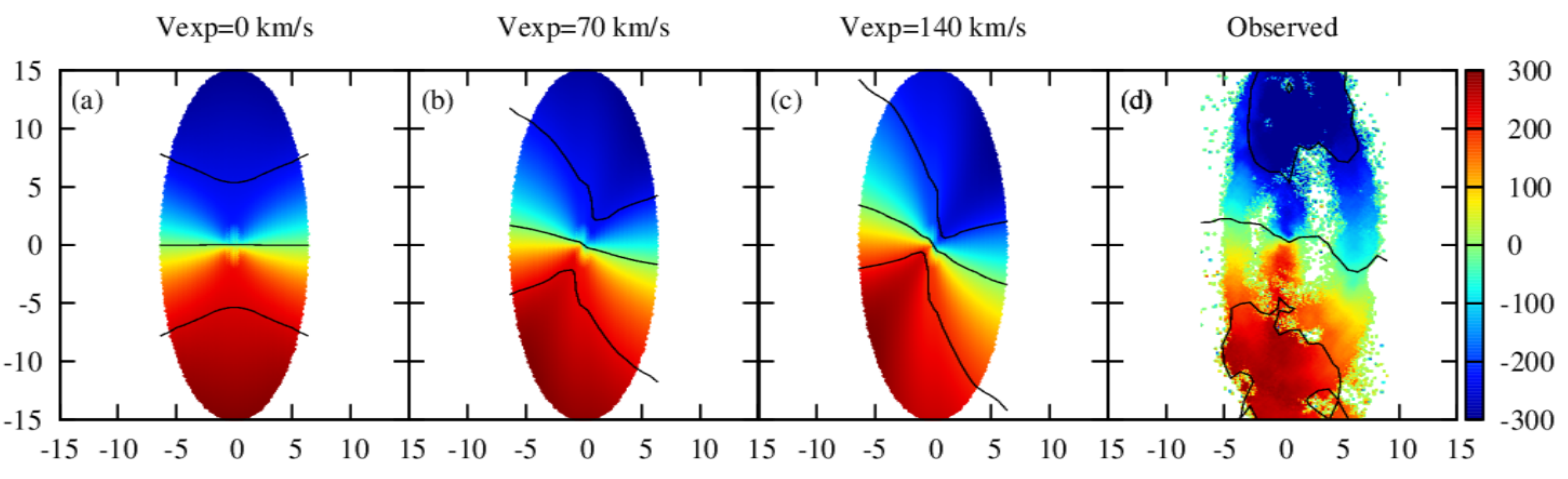}
\caption{\label{fig:model_velocities}Panels (a), (b) and (c) illustrate the line-of-sight velocity field of a toy model disk with a circular velocity of 320~km.s$^{-1}$. In panel (a) the disk is purely rotating. In panel (b) and (c), a component of radial expansion is added to the disk dynamics. Contours are shown at -200~km.s$^{-1}$, 0~km.s$^{-1}$, and 200~km.s$^{-1}$. The velocity field twists clockwise with increasing expansion velocity. In panel (d), the H$_\alpha$ velocity map of AM1354-250 is shown, rotated anticlockwise by 48$^\circ$ to match the position angle of the toy model. The observed twist in the velocity field is approximated by an expansion velocity of $\sim$ 70~km.s$^{-1}$. This suggests that AM1354-250 is dominated by rotation, with a weak additional component of expansion.}
\end{figure*}

We confirm our measured values of the expansion and rotation velocity of the ring by making velocity maps of toy model disks (see Figure \ref{fig:model_velocities}). Contours are shown at -200, 0, and 200~km.s$^{-1}$. In all the toy model disks, we fix the circular velocity to be 320~km.s$^{-1}$, but between panels we vary the radial expansion velocity of the toy model. As the expansion velocity is increased (moving from panel (a) towards panel (c)), an increasing twist in the velocity field appears. Panel (d) shows the observed H$_\alpha$ velocity map of AM1354-250, rotated so as its position angle matches that of the toy model. A similar twist in the velocity field of AM1354-250 can be seen. The strength of the twist is best approximated by the toy model with the expansion velocity of 70~km.s$^{-1}$ (panel (b)). This further suggests that the dynamics of the disk of AM1354-250 are dominated by rotation with an additional second order component of expansion. We note that the toy model is unable to tightly match other features of the observed velocity field, such as the exact shape of the 200~km.s$^{-1}$ and -200~km.s$^{-1}$ contour, partly due to regions where there are gaps in the H$_\alpha$ emission. However, it can be seen by eye that the velocity field is not symmetrical in the way seen in panel (a) in regions where the velocity is $>$200~km.s$^{-1}$ or $<$-200~km.s$^{-1}$, which supports the idea that some minor component of expansion is required on top of the rotation of the disk.

In comparison with \citet{Wallin1995}, our expansion velocity of $\sim$70\,km.s$^{-1}$ is slightly lower than their estimate of $\sim$95~km.s$^{-1}$ which was based on simple stellar population models. Given the intrinsic difficulties in deriving accurate ages using simple stellar population models however, we believe the agreement is actually very good, and the key point is that both independent methods prefer a ring that is strongly dominated by rotation. In fact, the expansion-to-rotation velocity of the ring is only $\sim$0.25. This is in contrast to Auriga's Wheel \citep{Conn2011} which was dominated by expansion (expansion-to-rotation ratio $>3$). Therefore the interaction that has occurred in AM1354-250 is comparatively minor and as such is driving a much weaker collisional ring. 

Supporting evidence for this picture comes from 2MASS Ks-band imaging where we can use the mass to light ratio to estimate the masses involved in the interaction. If we first consider the ring galaxy and assume a stellar mass-to-light ratio of 0.95 from Equation~\ref{eqn:mass}, we calculate that the total stellar mass of the ring galaxy within $4$~kpc (encompassing the bulge) to be $1.5 \times 10^{11} M_{\odot}$ this increases to $2.4 \times 10^{11} M_{\odot}$ within $11$~kpc (encompassing the ring as well). The collider stellar mass is determined to be $1.3 \times 10^{10} M_{\odot}$ within a radius of approximately $5$~kpc (encompassing the entire collider galaxy). Therefore the stellar mass ratio of the pre-collision spiral to the collider is between 1:12 and 1:20. Assuming an equal stellar-to-dark matter mass ratio for the two galaxies, the galaxies would maintain this total mass ratio of between 1:12 to 1:20. Thus the 2MASS Ks-band imaging and the expansion-to-rotation ratio of the ring support a scenario where a minor interaction generates the ring.

Determining the age of the collision at this stage of the analysis is limited since we are only able to use the most basic parameters of the system as we have just derived them. If we assume a constant expansion velocity since the collision, this would imply that the collision took place $140 \pm 12$~Myr ago. This assumption is reasonable to first order (for example, see Figure 11 of \citet{Smith2012}), however if we further assume that the derived relative velocity of the collider is a constant since the collision, then the collider would only be $23.9 \pm 4.2$~kpc from the disk. This is much lower than the distance derived from our observations. This shows that the assumption of constant relative velocity is incorrect, and suggests the relative velocity must have been higher previously. In fact, this is to be expected. As the collider climbs out of the potential well of the spiral, the relative velocity is expected to decrease rapidly. Rapid relative velocity evolution is also seen in modelling of other ring systems, e.g. Figure 7 of \citet{Smith2012}. To extend our understanding of the formation of AM1354-250 beyond measuring these basic parameters, we will conduct detailed numerical modelling. The parameters we have derived here will be used to constrain these models, which will be reported in a future paper \citep{simpaper}.

\section{Conclusion}
We have confirmed that AM1354-250 is a collisional ring galaxy, demonstrating the presence of an expanding star forming ring. There is no conclusive evidence of a second ring but the presence of a very large star forming region at the location of the expected second ring is indicative that such a scenario is possible.

We used 2MASS Ks-band imaging to derive mass estimates of $\sim$1.5$\times$10$^{11}$M$_{\odot}$ and $\sim$1.3$\times$$10^{10}$M$_{\odot}$ for ring and collider, respectively. The photometric position angle, ellipticity and surface brightness profiles were also determined from the 2MASS Ks-band image.

We used IFS observations from MUSE to synthesise high resolution $g$, $r$ and $i$ band images of the ring galaxy. We derived stellar velocity and velocity dispersion maps by fitting each spectrum in the MUSE data cube with a set of stellar templates. We then derived emission line intensity and velocity maps for H$\beta$, [OIII]$\lambda$5007, H$\alpha$ and [NII]$\lambda$6583 across the entire face of the ring galaxy. 

From these we determined the rotation curve of the galaxy in both gas and stars using {\sc Kinemetry}. We see a kinematically distinct, but aligned, stellar disk in the centre of the galaxy, whereas the rotation curve for the gas rises steadily, showing no kinematically distinct component.

To investigate the ionised gas in the galaxy we plotted each individual spaxel on a BPT diagram ([OIII]/H$\beta$ vs. [NII]/H$\alpha$). We see various ionisation mechanisms across the galaxy. Ionisation from star formation is dominant in the ring along with some evidence of shocked gas. In the central few kiloparsecs of the galaxy the line ratios reveal LINER-like emission with no evidence of AGN activity. We also see a large star forming region well outside the recognised disk of the galaxy. It is comparable to the large star forming knots in the ring itself and is approximately 2 kpc across.

Using the inferred rotation rate of the ring galaxy, from the H$\alpha$ velocity map, after correcting for inclination ($\sim$320\,km.s$^{-1}$) and fitting to the contour twists in the velocity maps, we have determined the expansion velocity of the ring to be $\sim$70\,km.s$^{-1}$. Assuming a constant expansion velocity since collision, we find that collision occurred $140 \pm 12$~Myr ago. This differs from the findings of \citet{Wallin1995} where a higher expansion velocity of $\sim$95\,km.s$^{-1}$ leads to a lower time since collision of less than 100 Myr ago, all derived from simple stellar population ages. Nevertheless, both approaches agree that the ring is heavily dominated by rotation as opposed to expansion, suggesting a fairly minor interaction has taken place. More detailed numerical modelling is needed to fit the interaction beyond the most basic parameters and these results will be presented in Candlish et al (in prep).

This dataset shows the full capabilities of the MUSE instrument and its ability to provide robust measurements of the various physical processes excited by interacting galaxies.

\acknowledgements

The Authors would like to thanks the Anonymous Referee for reviewing our paper and to thank the European Southern Observatory for awarding our project time during the Science Verification period of MUSE. RS acknowledges support from Brain Korea 21 Plus Program (21A20131500002) and the Doyak 1689 Grant (2014003730). RS also acknowledges support from the EC through an ERC grant StG-257720, and Fondecyt (project number 3120135). GNC acknowledges support of FONDECYT grant 3130480. 

This research has made use of data, software and/or web tools obtained from the High Energy Astrophysics Science Archive Research Center (HEASARC), a service of the Astrophysics Science Division at NASA/GSFC and of the Smithsonian Astrophysical Observatory's High Energy Astrophysics Division.

The Digitized Sky Surveys were produced at the Space Telescope Science Institute under U.S. Government grant NAG W-2166. The images of these surveys are based on photographic data obtained using the Oschin Schmidt Telescope on Palomar Mountain and the UK Schmidt Telescope. The plates were processed into the present compressed digital form with the permission of these institutions.

The National Geographic Society - Palomar Observatory Sky Atlas (POSS-I) was made by the California Institute of Technology with grants from the National Geographic Society.

The Second Palomar Observatory Sky Survey (POSS-II) was made by the California Institute of Technology with funds from the National Science Foundation, the National Geographic Society, the Sloan Foundation, the Samuel Oschin Foundation, and the Eastman Kodak Corporation.

The Oschin Schmidt Telescope is operated by the California Institute of Technology and Palomar Observatory.

The UK Schmidt Telescope was operated by the Royal Observatory Edinburgh, with funding from the UK Science and Engineering Research Council (later the UK Particle Physics and Astronomy Research Council), until 1988 June, and thereafter by the Anglo-Australian Observatory. The blue plates of the southern Sky Atlas and its Equatorial Extension (together known as the SERC-J), as well as the Equatorial Red (ER), and the Second Epoch [red] Survey (SES) were all taken with the UK Schmidt.

All data are subject to the copyright given in the copyright summary. Copyright information specific to individual plates is provided in the downloaded FITS headers.

Supplemental funding for sky-survey work at the ST ScI is provided by the European Southern Observatory.

\end{document}